\documentclass[twocolumn,10pt,prx,floatfix]{revtex4-1}

\input{glyphtounicode}
\usepackage[T1]{fontenc}
\usepackage[utf8]{inputenc}
\usepackage[english]{babel}
\usepackage{amsmath}  
\usepackage{amsfonts} 
\usepackage{graphicx} 
\usepackage[pdftex,bookmarks=true,bookmarksopen,bookmarksnumbered,
                colorlinks,
                linkcolor=blue,
                citecolor=blue]{hyperref}
\graphicspath{ {./} }
\usepackage{natbib}
\usepackage{array}
\usepackage[table]{xcolor}
\usepackage{ulem}
\usepackage{xcolor}
\usepackage{braket}
\usepackage{booktabs}
\usepackage{multirow}
\usepackage{caption}
\usepackage{float}

\newcommand{\aref}[1]{\hyperref[#1]{Appendix~\ref*{#1}}}

\begin{document}

\captionsetup[table]{name={TABLE},labelsep=period,justification=raggedright,font=small}
\captionsetup[figure]{name={FIG.},labelsep=period,justification=raggedright,font=small}
\renewcommand{\equationautorefname}{Eq.}
\renewcommand{\figureautorefname}{Fig.}
\renewcommand*{\sectionautorefname}{Sec.}

\title{Accessing the Full Capabilities of Filter Functions: A Tool for Detailed Noise and Control Susceptibility Analysis}

\author{Ingvild Hansen$^1$}
\author{Amanda E. Seedhouse$^1$}
\author{Andre Saraiva$^{1,2}$}
\author{Andrew S. Dzurak$^{1,2}$}
\author{Chih Hwan Yang$^{1,2}$}
\affiliation{$^1$School of Electrical Engineering and Telecommunications, The University of New South Wales, Sydney, NSW 2052, Australia}
\affiliation{$^2$Diraq, Sydney, New South Wales, Australia}

\date{\today}

\begin{abstract}

The filter function formalism from quantum control theory is typically used to determine the noise susceptibility of pulse sequences by looking at the overlap between the filter function of the sequence and the noise power spectral density. Importantly, the square modulus of the filter function is used for this method, hence directional and phase information is lost. In this work, we take advantage of the full filter function including directional and phase information. By decomposing the filter function with phase preservation before taking the modulus, we are able to consider the contributions to $x$-, $y$- and $z$-rotation separately. Continuously driven systems provide noise protection in the form of dynamical decoupling by cancelling low-frequency noise, however, generating control pulses synchronously with an arbitrary driving field is not trivial. Using the decomposed filter function we look at the controllability of a system under arbitrary driving fields, as well as the noise susceptibility, and also relate the filter function to the geometric formalism. 
\end{abstract}

\pacs{}

\maketitle

\section{Introduction}
\label{sec:intro}

One of the biggest hurdles in realising a large-scale quantum computer is mitigating decoherence caused by noise in the qubit environment. Qubits are notoriously sensitive to noise originating from the material stack, control instruments, etc. \cite{Saraiva2022,chan2018}. There are many means to reduce the effect of noise sources in qubit devices such as isotropic purification of the host material \cite{Itoh2014} and feedback protocols \cite{Yang2019,Vep2022,philips2022}, however the residual noise still limits qubit performance. 

Different qubit modalities are plagued by different noise frequency distributions. Qubits in the solid state, for example, are typically dominated by 1/$f$ noise \cite{chan2018,Paladino2014}. One method to tackle this type of noise is through continuous driving. Driven two-level systems, often referred to as microwave dressed systems \cite{Baur2009,laucht2017dressed, Seedhouse2021}, are continuously decoupled from environmental noise. Moreover, by tailoring the amplitude modulation of the driving field one can achieve higher order noise protection \cite{Hansen2021}. This novel control strategy is compatible with global control \cite{kane1998,Seedhouse2021}, which is promising for scalability. Single qubit control is implemented by applying an additional local control pulse, to dynamically control the individual qubit frequency,
on top of the continuous global microwave drive. However, working out the required waveform of the local control pulses to be applied synchronously with the continuous driving field is not trivial.

Here we develop a generalised method to find the waveform of the local control pulses required for two-axis control of a two-level system driven by an arbitrary global driving field. The global driving field decouples the system from low frequency noise, whereas the local control pulses allow for single qubit addressability. We find the local control pulses by decomposing the filter function calculated from Magnus expansion series and including phase information. We show how this is related to the geometric formalism \cite{Zeng2019,Barnes_2022} and how the accumulated effect of noise or control pulses is expressed in a 3D space representation. This work expands on similar methods \cite{Green2013,Hangleiter2021,Cerfontaine2021} that were previously developed for noise analysis, with tools for a more general analysis of the control of driven systems. 

The method developed in this work is specifically important in the context of driven protocols \cite{Hansen2021,Seedhouse2021,hansen2022} in order to understand the origin of the local control pulses and the noise cancelling properties of the global driving fields.

\subsection*{Calculation of the complex filter function}
\label{sec:calc}

\begin{figure*}
    \centering
    \includegraphics[width=0.7\linewidth]{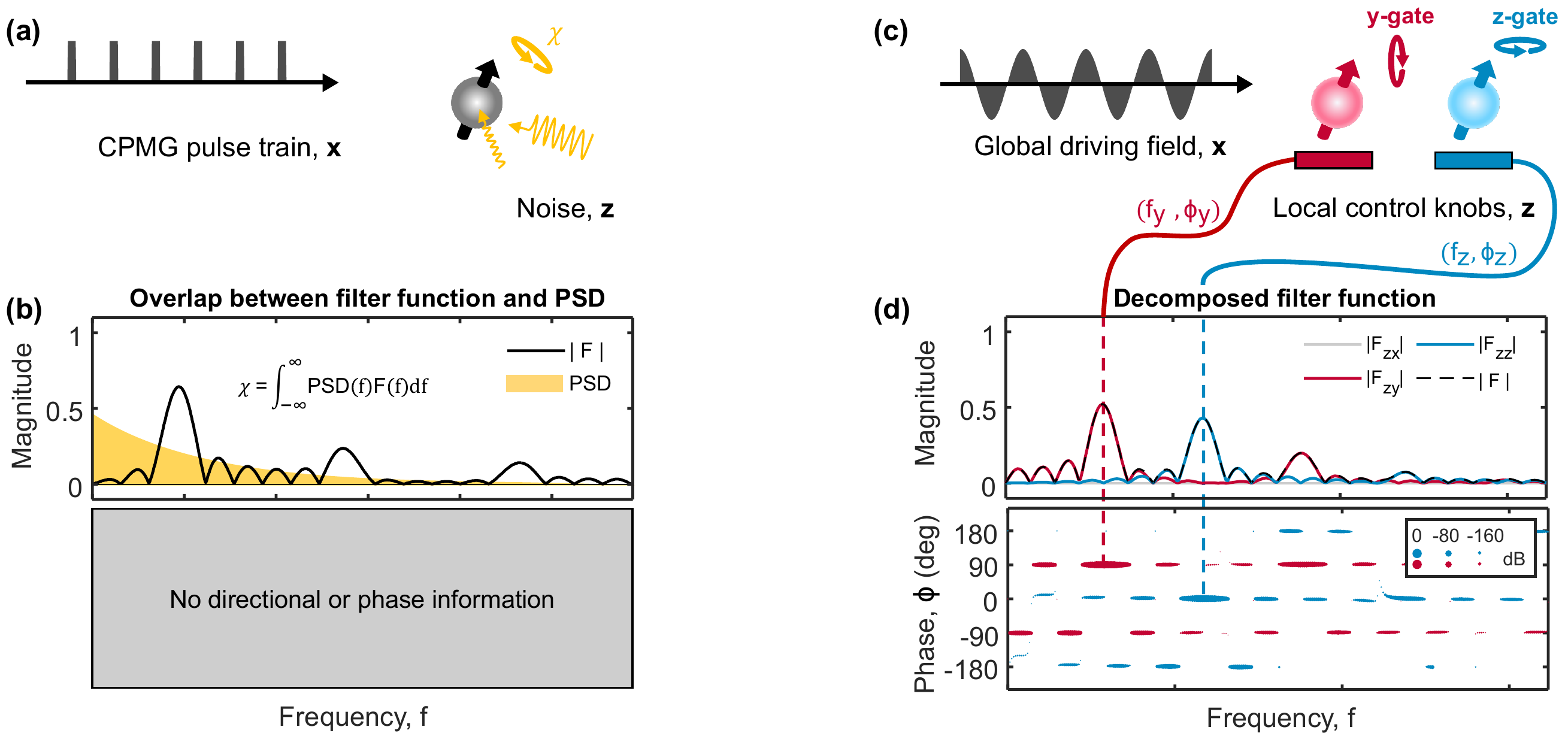}
    \caption{{\bfseries{Common use of filter function for noise susceptibility and improved method for control susceptibility.}} (a) A CPMG pulse train produces a filter function (b) which overlap with the power spectral density results in some noise rotation magnitude $\chi$. (c) For a driven system the decomposed filter function gives access to information about direction and phase (d), which can be used as two-axis control parameters to be applied synchronous with the global field through local control knobs.
    }
\label{fig:controlfilter}
\end{figure*}

A driven qubit subject to a single quasistatic noise source can be described by the rotating frame Hamiltonian
\begin{equation}
\begin{split}
        H_{{\rm{r}}}(t)&=H_{\rm drive}(t)+\delta{H}\\
        &=\Omega_{{\rm{RI}}}(t)\sigma_x+\Omega_{{\rm{RQ}}}(t)\sigma_y
        +{\delta\beta}{\sigma_i}.
\end{split}
\end{equation}
Here, $\Omega_{{\rm{RI/Q}}}(t)$ describes the envelope of arbitrary driving fields along perpendicular axes, $\sigma_n$ ($n=x,y,z$) the Pauli matrices  and $\delta\beta$ is a small perturbation term. We will from now on assume a driving field along $x$, and a perturbation along $z$ ($\sigma_i=\sigma_z$). One can think of the perturbation as either stochastic noise or an intentionally applied control pulse, which can be seen in \autoref{fig:controlfilter}, where the spin in (a) has a noise perturbation and the other two spins in (c) have control perturbations. We will start out by treating $\delta\beta$ as noise, and come back to control later.

It is useful to move into an interaction picture set by the global driving field, in order to isolate the effect of the perturbation in a noise-propagator 
\begin{equation}
\begin{split}
    H_z^{\rm{int}}(t)&={\delta\beta{U}_{\rm drive}^{\dagger}(t)\sigma_z{U}_{\rm drive}(t)}\\
    &=\delta\beta\sum_{j=x,y,z}{{R}_{j}(t)\sigma_j},
\end{split}
\label{eq:hi}
\end{equation}
%



where the rotation matrix is given by \cite{Green2013}
\begin{equation}
    R_{j}(t)={\rm{Tr}}({U}_{\rm drive}^{\dagger}(t)\sigma_z{U}_{\rm drive}(t)\sigma_j)/2.
\end{equation}
Here, $U_{\rm{drive}}$ is the time evolution operator derived from $H_{\rm{drive}}$. By integrating $H_z^{\rm{int}}(t)$ over time, we find the accumulated effect from the perturbation in the frame of the driving field, described in first order Magnus expansion
\begin{equation}
    A_{1z}(t)=\frac{1}{\delta\beta}\int_{0}^{t}{H_z^{\rm{int}}(t)dt}.
    \label{eq:A1}
\end{equation}
 Here, $z$ represents the perturbation axis. Higher order terms of the Magnus expansion can be included for a more accurate representation of the accumulated effect of the noise. The perturbation magnitude $\delta\beta$ is in general limited to small values compared to the terms in $H_{\rm{drive}}$ to avoid higher order terms. That is, we are restricted to the weak noise (or control) regime.

\autoref{eq:A1} equals zero if the global driving field cancels out the effect of the perturbation to the first order. On the other hand, if the perturbation introduces rotation about $\sigma_k$ after time $T$, $A_{1z}(T)$ is proportional to $\sigma_k$. \autoref{eq:A1} is used in the geometric formalism in order to describe cancellation of quasistatic noise using 3D space curves \cite{Zeng2019}. Details about the connection between filter functions and the geometric formalism are discussed later.
 
 By multiplying the integrand in \autoref{eq:A1} with a complex exponential function, we look at the effect of the perturbation at different frequencies and phases. That is, we replace the quasistatic perturbation term $\delta\beta$ with an a.c. complex term $\delta\beta\rm{e}^{i2\pi{f}t}$. This results in a generalized form of the interaction picture and what we call the first order filter function \cite{Green2013} of the driving field 

\begin{equation}
   F_z(f,t)=\frac{1}{\delta\beta{t}}\int_{0}^{t}{H_z^{\rm{int}}(\tau){\rm{e}}^{i2\pi{f(\tau-t/2)}}d\tau}.
   \label{eq:filter0}
\end{equation}

Here, the complex exponential represents a perturbation with defined frequency $f$ applied for a time $t$ along axis $z$. By using a complex exponential we can assess the phase as well as frequency sensitivity. Using a symmetry argument, we subtract the total time divided by two in the complex exponential. This effectively sets the phase of the perturbation from the centre of the sequence, which simplifies the phase information for plotting/presentation purposes. The filter function is normalised with time so that the magnitude corresponds to the rotation efficiency and by increasing the total time of the driving, the filter bandwidth becomes narrower as expected from Fourier analysis. 


%
\begin{figure*}[hbt!]
    \centering
    \includegraphics[width=0.85\textwidth]{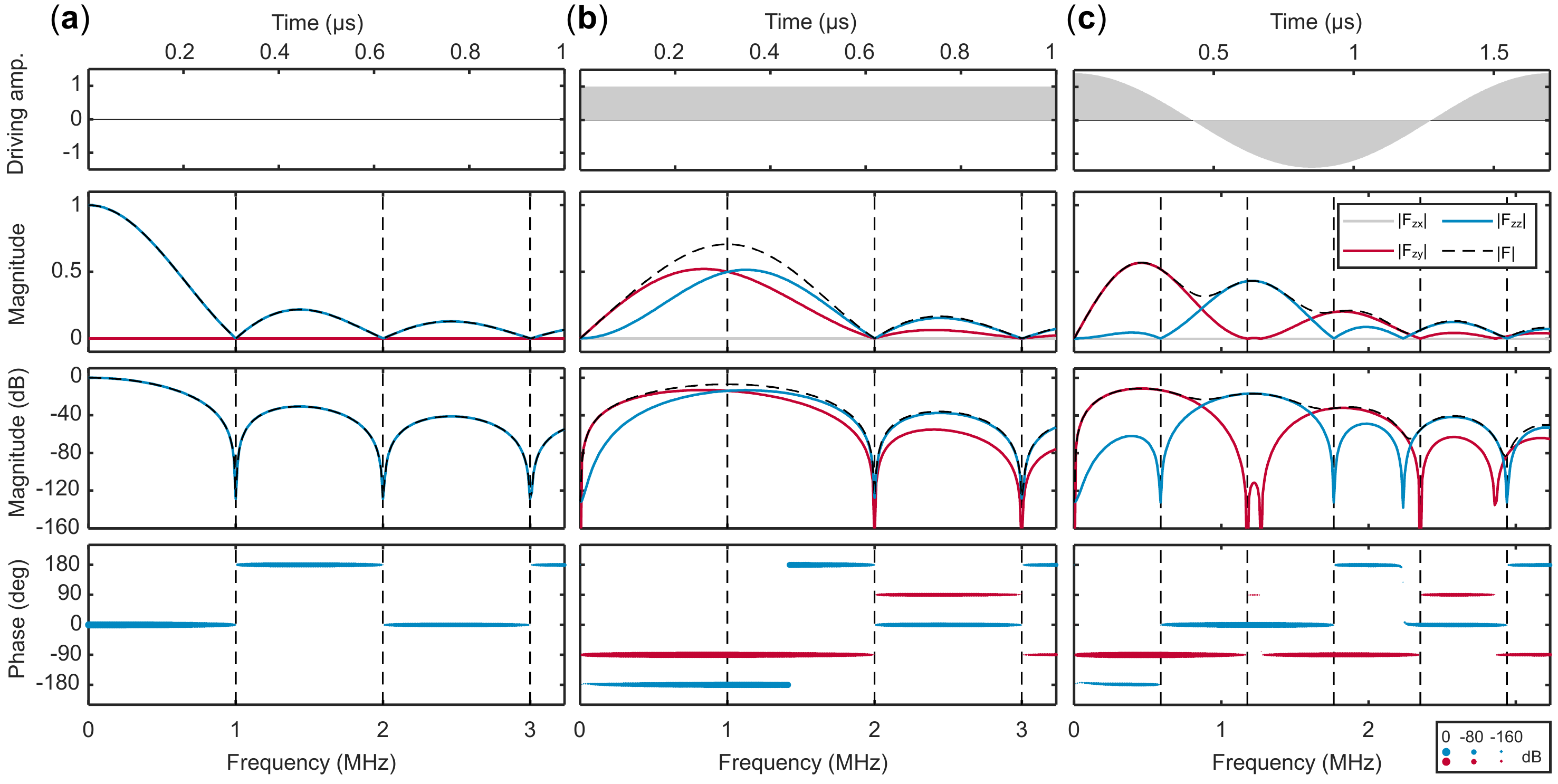}
    \caption{{\bfseries{Examples of the full filter function.}} Driving field over a time $T$, filter function magnitude (unitless and dB) and phase for (a) a bare system, (b) a dressed system and (c) a cosine modulated dressed system when assuming noise on $z$-axis. Multiples of $1/T$ are shown with vertical dotted lines.
    }
    \label{fig:filter_principle}
\end{figure*}
\begin{figure*}[hbt!]
    \centering
    \includegraphics[width=0.85\textwidth]{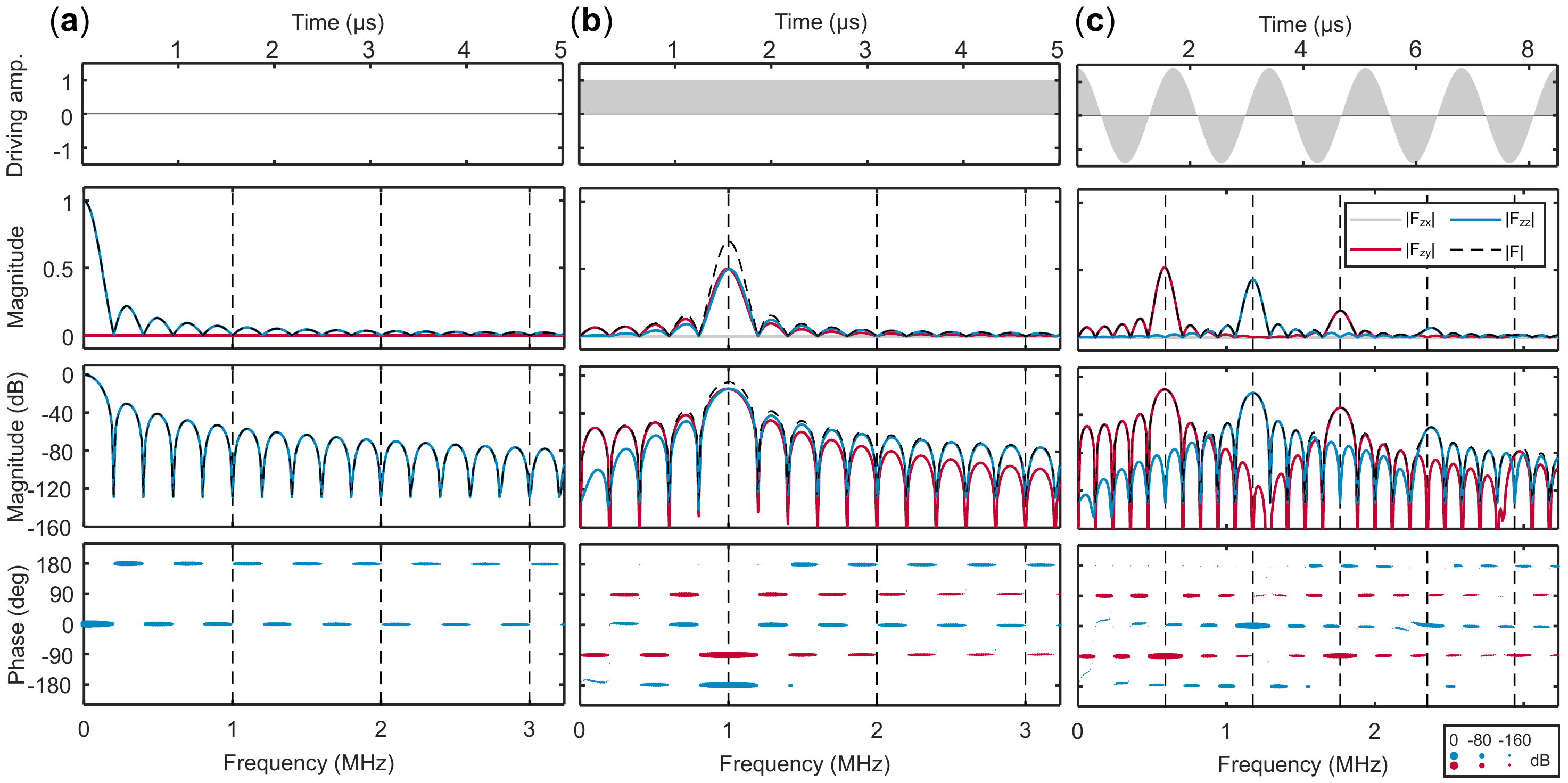}
    \caption{{\bfseries{Multi-period filter function.}} The same examples as in \autoref{fig:filter_principle} but with a time corresponding to five periods. (a) Bare system, (b) dressed system and (c) cosine modulated dressed system.
    }
    \label{fig:filter_principle2}
\end{figure*}
%


In \autoref{eq:filter0}, $F_z(f,t)$ is a unitless $2\times2$ complex matrix that can be decomposed by taking the trace with the product of the Pauli matrices according to $F_{ij}(f,t)={\rm{Tr}}(F_i(f,t)\sigma_{j})$. Here, $j$ represents the Pauli component of $F_i$. The absolute value $|F_{ij}|$ represents the gain of the filter function at different frequencies along axis $j$ when the perturbation is along axis $i$. By using the Euler formula arbitrary phase response can be found as well. 


%
\begin{figure*}[hbt!]
    \centering
    \includegraphics[width=0.85\textwidth]{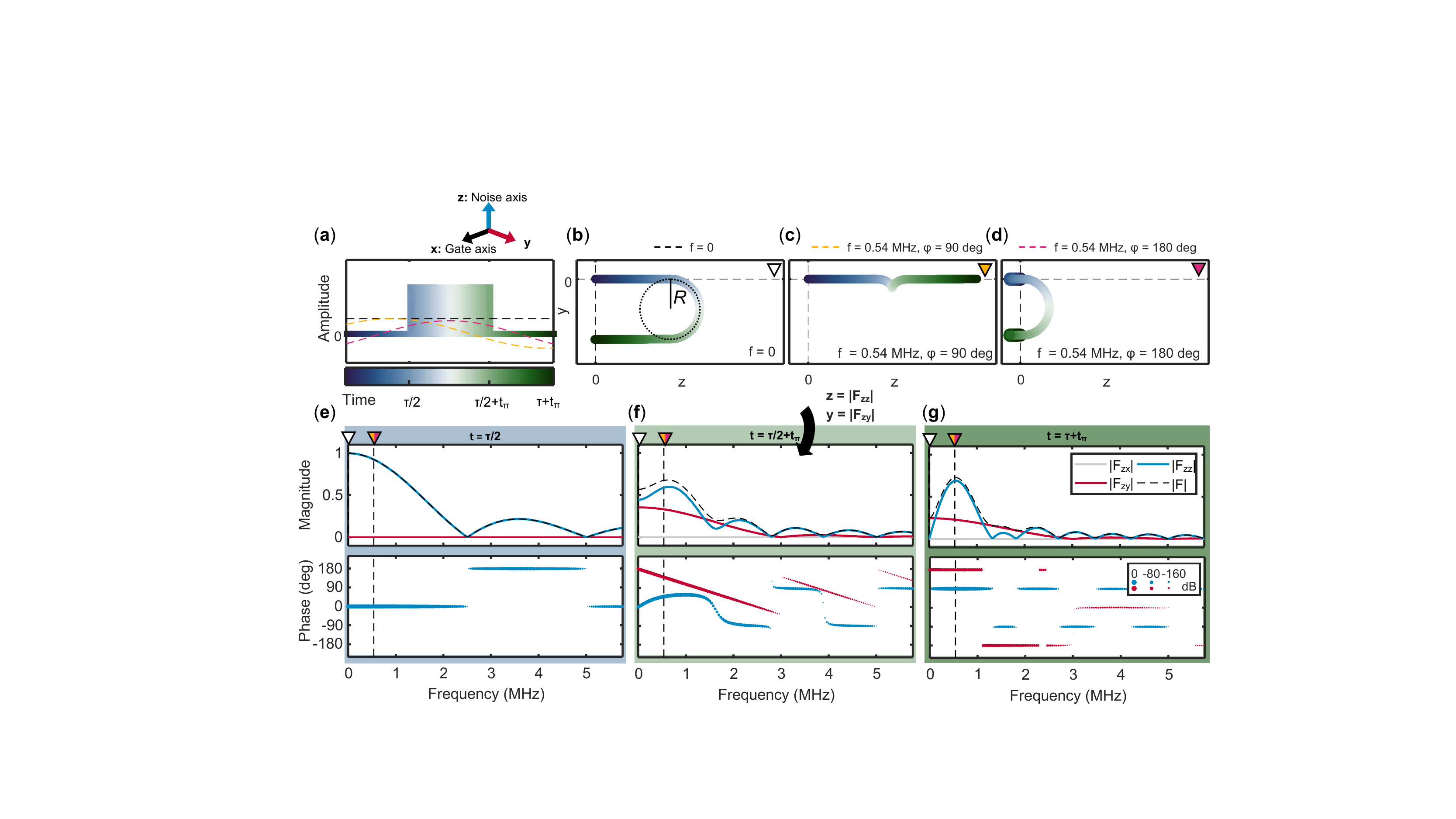}
    \caption{{\bfseries{Noise susceptibility from geometric formalism and filter function.}} (a) Hahn echo sequence for $\pi$-gate on $x$-axis with three different noise implementations. (b-d) The three corresponding geometric space curves. (e-g) Filter function at three specific time steps when perturbation axis is along $z$. Here $\tau/2=0.4\,\mu$s and $t_{\pi}=0.5\,\mu$s. 
    }
    \label{fig:filtergeo}
\end{figure*}
%


The method described above is a more complete way of calculating the filter function compared to the conventional one \cite{Biercuk2009,Biercuk2011,Frey2017,Kofman2001,Kofman2004}, where the squared absolute value of the Fourier transform of the time-domain filter function/switching function is used
\begin{equation}
   F(f,t)=|\tilde{y}(ft)|^2.
    \label{eq:trad}
\end{equation}
The switching function $y(t)$ toggles between $+{1}$ and $-{1}$ for every applied $\pi$-pulse between free precession periods. For CPMG spectroscopy, the overlap between the power spectral density and the filter function is then used to evaluate the noise sensitivity [PSD (yellow) and $|F|$ (dotted line) in \autoref{fig:controlfilter}(b)]. However, by calculating the filter function in this simplified way, all directional and phase information is lost.

We study next a few examples to help interpret the full information from the complex filter function. In \autoref{fig:filter_principle} the driving fields and filter functions for a bare, dressed and modulated dressed system (according to the SMART protocol \cite{hansen2022}) are shown. For the dressed example we use an amplitude corresponding to $1$\,MHz Rabi oscillations and the same power for the modulated dressed example. However, any frequency scale is applicable. Here the three components of the decomposed filter function are shown together with the square root of the sum of squares. The latter represents the simplified version (\autoref{eq:trad}). 
The magnitude of each component represents the effective rotation about the respective axis and the phase corresponds to the maximum magnitude at any frequency. 

The bare system in \autoref{fig:filter_principle}(a) is sensitive to d.c. noise with decreasing filter gain for higher frequencies. Low-frequency noise causes rotation about $z$ which can be seen from the non-zero value of $|F_{zz}|$ (blue trace). The bare system filter function has zero magnitude at multiples of $1/T$ (dotted vertical lines) where $T$ is the total duration. These are the cases where the noise effectively results in an identity gate (e.g. $2\pi$ phase accumulation). The bare system can in fact not be seen as a perturbation problem, since $\delta\beta\sigma_z$ is the only term in the Hamiltonian.

The dressed system in \autoref{fig:filter_principle}(b), on the other hand, is insensitive to d.c. noise and mainly responds to $1/T$ as indicated by the $|F_{zz}|$ and $|F_{zy}|$ peaks at $1$\,MHz. This frequency noise causes $z$-rotation or $y$-rotation depending on the phase of the noise.

The cosine modulated dressed case in \autoref{fig:filter_principle}(c) is also insensitive to d.c. noise, but responds to $1/T$ and its multiples with decreasing filter gain for increasing frequency. In this case the peaks of $|F_{zy}|$ and $|F_{zz}|$ are separated both in frequency and phase, with corresponding frequencies $1/T=\sim0.6$\,MHz and $2/T=\sim$1.2\,MHz and phases $-90$\,deg and $0$\,deg. In \autoref{fig:filter_principle2} the same driving fields are used but with five times longer duration, showing narrower bandwidth of the filter functions.

\subsection*{From filter functions to the geometric formalism}
\label{sec:from}

The geometric formalism is the particular case of $f=0$ filter function \cite{Zeng2019}. Here we expand the geometric formalism to include noise with frequency $f$ to relate it to the entire filter function. Now we can describe the filter function and the space curve with the same equation (\autoref{eq:filter0}). We will see in the next example that the filter function is assigned a space curve for each frequency and phase.

We will use the example of the Hahn echo sequence to explain the connection between the filter function and the geometric formalism. We use a $\pi$-pulse on the $x$-axis with a finite duration, as illustrated in \autoref{fig:filtergeo}(a), where three different noise frequency cases are considered with space curves in (b-d) and three different times are looked at for the filter function in (e-g) ($\tau/2, \tau/2+t_{\pi}$ and $\tau+t_{\pi}$). The connection between the filter function components and the 3D space curve in \autoref{fig:filtergeo} is given by $|F_{zz}|=z$, $|F_{zy}|=y$ and $|F_{zx}|=x$ for a given time. 

For the d.c. noise case in \autoref{fig:filtergeo}(b) the system is under free precession for $t\leq{\tau/2}$. This equates to a space curve moving from the origin to an increasing $z$-values with time. The filter function accumulates a corresponding $|F_{zz}|$ value [blue trace at $f=0$ in \autoref{fig:filtergeo}(e)]. During the $\pi$-pulse the space curve makes a semi-circle in the $zy$-plane where the radius $R$ increases with the inverse of the Rabi frequency. This equates to $|F_{zy}|$ accumulating in \autoref{fig:filtergeo}(f) (red trace at $f=0$). After the final free precession the space curve is back at $z=0$, but now the value of $y$ is $2 \times R$. For the filter function we see in \autoref{fig:filtergeo}(g) that $|F_{zz}|$ is zero whereas $|F_{zy}|$ is not (blue and red trace at $f=0$, respectively).

For the noise cases with $f=0.54$\,MHz the system is under free precession at $t\leq{\tau/2}$, and the space curve moves towards positive or negative $z$-values depending on the phase/sign of the noise [\autoref{fig:filtergeo}(c-d)]. The $|F_{zz}|$ filter function magnitude at $f=0.54$\,MHz in \autoref{fig:filtergeo}(d) is lower than for the d.c. case due to lower noise power. During the $\pi$-pulse for the case with $f=0.54$\,MHz and $\phi=90$\,deg the phase shift caused by the $\pi$-pulse coincides with the change of sign of the noise [\autoref{fig:filtergeo}(c)]. Therefore, the space curve continues to accumulate $z$-value and the $\pi$-pulse enhances the noise instead of cancelling it out. The case with $f=0.54$\,MHz and $\phi=180$\,deg, on the other hand, makes an approximate semi-circle and comes back to $z=0$ but with a finite $y$-value. From the filter function data in \autoref{fig:filtergeo}(g) this can be seen from the $|F_{zz}|$ peak at $f=0.54$\,MHz when $\phi=90$\,deg and a non-zero $|F_{zy}|$ values at the same frequency when $\phi=180$\,deg.

\subsection*{Controllability using decomposed filter functions}
\label{sec:ctrl}

The geometric formalism and the filter function are first of all used to characterise noise performance. However, both can also be used to describe control. We will now treat the perturbation $\delta\beta{e^{i2\pi{ft}}}$ as control (to be applied synchronous with the global driving field). The noise-propagator in \autoref{eq:hi} can then be seen as a control-propagator instead. In other words, coherent noise with defined frequency and phase is equivalent to control pulses. 

Controllability for an arbitrary global driving field is found by locating the peaks in the filter function and extracting the frequency and phase. By applying control pulses with these parameters synchronously with the global field, the desired rotation occurs. In the geometric formalism, this corresponds to a space curve ending up somewhere along the axis of rotation. The rotation magnitude is represented by the filter function gain and the displacement from the origin of the space curve.

 This new gate representation, using the filter function and the geometric formalism, is useful to look at the effective rotation axis for a given driving field as a function of frequency, phase and duration. Hence, the filter function and geometric formalism are not only useful for stochastic quasistatic perturbation in the form of noise, but also intentional a.c. control perturbations. The gate representation differs from the one in \cite{Zeng2019} where the control pulse is included in $H_{{\rm{drive}}}(t)$ and the gate is represented by the orientation of the final tangent vector relative to the initial one.

\begin{figure}[hbt!]
\centering
    \includegraphics[width=0.75\linewidth]{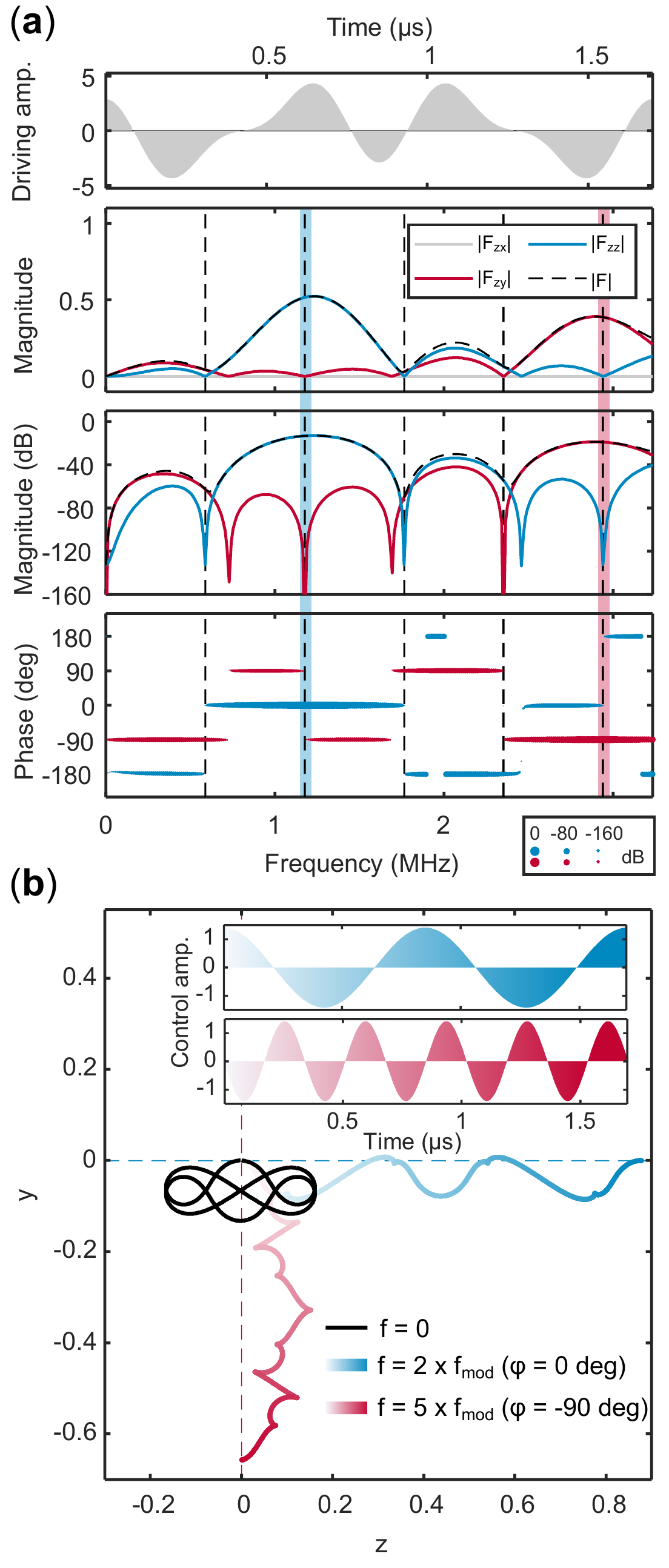}
   \caption{{\bfseries{Gate representation with filter function and geometric formalism.}} (a) The driving field constructed from the weighted sum of three harmonics and the corresponding filter function, (b) controls and space curves. The space curves for three different frequencies are shown. Two-axis controllability is found with the second and fifth harmonic.
   }
   \label{fig:controllability}
\end{figure}

Finally we will show an example of how to use the filter function to find two-axis control of a complicated three-harmonic global driving field [\autoref{fig:controllability}(a)]. This specific driving field provides as high as six order noise cancellation (see Appendix A for more details). The filter function peaks are located at the 2nd and 5th harmonic (vertical dotted lines at $\sim$1.2\,MHz and $\sim$3\,MHz) with phase $0$\,deg ($z$) and $-90$\,deg ($y$), respectively. The controls are also represented using the corresponding space curves in \autoref{fig:controllability}(b). We find that there are several frequency options that enable two-axis control, however, for quantum computation we want fast gates in order to do maximal gate repetitions before decohering and we therefore choose the most efficient gate implementations with higher filter gain. Two-axis control of the two dressed systems in \autoref{fig:filter_principle2}(b-c) is also found by simply locating the filter function peaks. These have been demonstrated experimentally in \cite{laucht2017dressed,hansen2022}.

\subsection*{Conclusion}
\label{sec:conc}
\vspace{-0.5cm}
In conclusion we have developed a generalised method to extract the local control pulses for systems driven by arbitrary global fields. In designing the global driving field both the qubit noise and control susceptibility and the power spectral density must be kept in mind. This allows for controllable arrays of qubits robust against low-frequency noise throughout the entire computation with control frequencies in strategic bands. The method is also useful for noise analysis including directional and phase information.

\section*{Acknowledgments}

We acknowledge support from the Australian Research Council (FL190100167 and CE170100012) and the US Army Research Office (W911NF-17-1-0198). The views and conclusions contained in this document are those of the authors and should not be interpreted as representing the official policies, either expressed or implied, of the Army Research Office or the U.S. Government. The U.S. Government is authorized to reproduce and distribute reprints for Government purposes notwithstanding any copyright notation herein. I.H and A.E.S acknowledge support from Sydney Quantum Academy.

\appendix

\section{Higher order filter functions}
\label{app:higher}

\begin{figure*}[hbt!]
    \centering
    \includegraphics[width=0.6\textwidth]{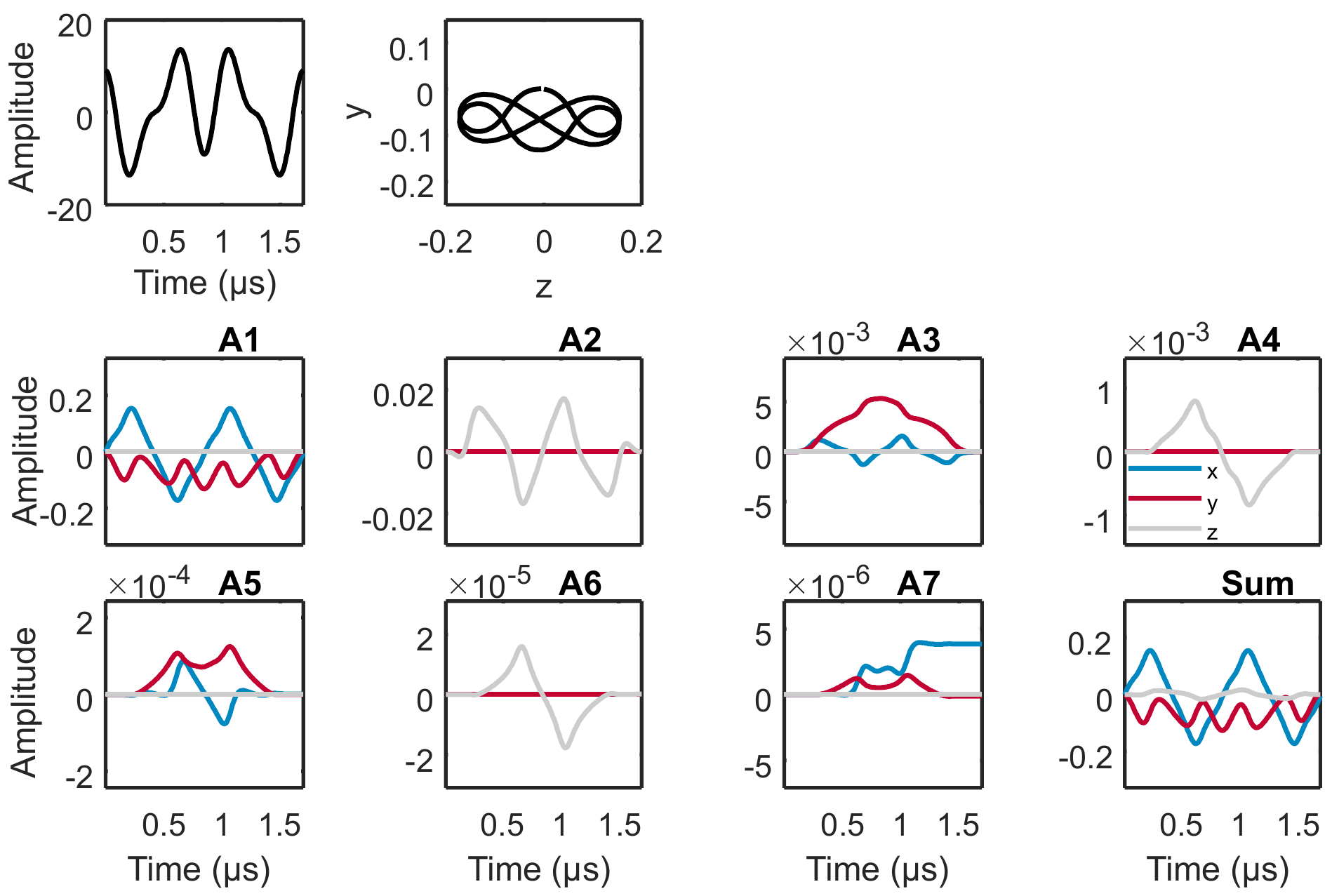}
    \caption{{\bfseries{Higher order Magnus expansion terms.}} Driving field, space curve and higher order Magnus expansion terms for one period of a driving field consisting of three harmonics with quasistatic noise assumption, showing noise cancellation up to 6th order. The noise magnitude equals the driving amplitude. 
    }
    \label{fig:threeHarmonic}
\end{figure*}

In \autoref{eq:filter0} the filter function is calculated using the first term of the Magnus expansion series. However, we can truncate the series at higher order and this will given a more accurate representation of the noise/control susceptibility of the system. Higher order terms of the Magnus expansion series are calculated with iterative integrals of right-nested commutators \cite{Arnal2018}. The full Magnus expansion when considering only $i$-axis noise is given by
\begin{equation}
    \begin{split}
        A_{i}(t)=\delta_{i}A_{1i}(t)+\delta_{i}^2 A_{2i}(t)+\delta_{i}^3 A_{3i}(t)+...
    \end{split}
\end{equation}
Using GRAPE we find that by combining three harmonics for the driving field we are able to cancel out quasistatic noise up to the sixth order. The parameters for the driving field are given here
\begin{equation}
\begin{split}
    H(t)=\Omega\bigg(h_1\cos(2\pi{ft})+h_3\cos(6\pi{ft})+h_5\cos(10\pi{ft})\bigg)
\end{split}
\end{equation}
\begin{equation}
    \begin{split}
    h_1=\cos(\gamma)\cos(\delta) \\
    h_2=\cos(\gamma)\sin(\delta) \\
    h_3=\sin(\gamma) \hspace{0.95cm}\\
    \end{split} 
\end{equation}

with $\Omega=-2.57453, \gamma=-0.49001,\delta=-1.04785$. The seven first terms of the three-harmonic drive for quasistatic noise are shown in \autoref{fig:threeHarmonic}, where the magnitude of the higher order terms are seen to decrease exponentially. This can be seen from the sum which closely resembles the first term. The first non-zero term represents the $T_1$ axis of the driven system, in this case $\sigma_x$. 

\onecolumngrid

\newpage
\twocolumngrid

\newpage

\bibliography{bib1}

\begin{thebibliography}{24}%
\makeatletter
\providecommand \@ifxundefined [1]{%
 \@ifx{#1\undefined}
}%
\providecommand \@ifnum [1]{%
 \ifnum #1\expandafter \@firstoftwo
 \else \expandafter \@secondoftwo
 \fi
}%
\providecommand \@ifx [1]{%
 \ifx #1\expandafter \@firstoftwo
 \else \expandafter \@secondoftwo
 \fi
}%
\providecommand \natexlab [1]{#1}%
\providecommand \enquote  [1]{``#1''}%
\providecommand \bibnamefont  [1]{#1}%
\providecommand \bibfnamefont [1]{#1}%
\providecommand \citenamefont [1]{#1}%
\providecommand \href@noop [0]{\@secondoftwo}%
\providecommand \href [0]{\begingroup \@sanitize@url \@href}%
\providecommand \@href[1]{\@@startlink{#1}\@@href}%
\providecommand \@@href[1]{\endgroup#1\@@endlink}%
\providecommand \@sanitize@url [0]{\catcode `\\12\catcode `\$12\catcode
  `\&12\catcode `\#12\catcode `\^12\catcode `\_12\catcode `\%12\relax}%
\providecommand \@@startlink[1]{}%
\providecommand \@@endlink[0]{}%
\providecommand \url  [0]{\begingroup\@sanitize@url \@url }%
\providecommand \@url [1]{\endgroup\@href {#1}{\urlprefix }}%
\providecommand \urlprefix  [0]{URL }%
\providecommand \Eprint [0]{\href }%
\providecommand \doibase [0]{http://dx.doi.org/}%
\providecommand \selectlanguage [0]{\@gobble}%
\providecommand \bibinfo  [0]{\@secondoftwo}%
\providecommand \bibfield  [0]{\@secondoftwo}%
\providecommand \translation [1]{[#1]}%
\providecommand \BibitemOpen [0]{}%
\providecommand \bibitemStop [0]{}%
\providecommand \bibitemNoStop [0]{.\EOS\space}%
\providecommand \EOS [0]{\spacefactor3000\relax}%
\providecommand \BibitemShut  [1]{\csname bibitem#1\endcsname}%
\let\auto@bib@innerbib\@empty
\bibitem [{\citenamefont {Saraiva}\ \emph {et~al.}(2022)\citenamefont
  {Saraiva}, \citenamefont {Lim}, \citenamefont {Yang}, \citenamefont {Escott},
  \citenamefont {Laucht},\ and\ \citenamefont {Dzurak}}]{Saraiva2022}%
  \BibitemOpen
  \bibfield  {author} {\bibinfo {author} {\bibfnamefont {A.}~\bibnamefont
  {Saraiva}}, \bibinfo {author} {\bibfnamefont {W.~H.}\ \bibnamefont {Lim}},
  \bibinfo {author} {\bibfnamefont {C.~H.}\ \bibnamefont {Yang}}, \bibinfo
  {author} {\bibfnamefont {C.~C.}\ \bibnamefont {Escott}}, \bibinfo {author}
  {\bibfnamefont {A.}~\bibnamefont {Laucht}}, \ and\ \bibinfo {author}
  {\bibfnamefont {A.~S.}\ \bibnamefont {Dzurak}},\ }\href {\doibase
  10.1002/adfm.202105488} {\bibfield  {journal} {\bibinfo  {journal} {Advanced
  Functional Materials}\ }\textbf {\bibinfo {volume} {32}},\ \bibinfo {pages}
  {2105488} (\bibinfo {year} {2022})}\BibitemShut {NoStop}%
\bibitem [{\citenamefont {Chan}\ \emph {et~al.}(2018)\citenamefont {Chan},
  \citenamefont {Huang}, \citenamefont {Yang}, \citenamefont {Hwang},
  \citenamefont {Hensen}, \citenamefont {Tanttu}, \citenamefont {Hudson},
  \citenamefont {Itoh}, \citenamefont {Laucht}, \citenamefont {Morello},\ and\
  \citenamefont {Dzurak}}]{chan2018}%
  \BibitemOpen
  \bibfield  {author} {\bibinfo {author} {\bibfnamefont {K.~W.}\ \bibnamefont
  {Chan}}, \bibinfo {author} {\bibfnamefont {W.}~\bibnamefont {Huang}},
  \bibinfo {author} {\bibfnamefont {C.~H.}\ \bibnamefont {Yang}}, \bibinfo
  {author} {\bibfnamefont {J.~C.~C.}\ \bibnamefont {Hwang}}, \bibinfo {author}
  {\bibfnamefont {B.}~\bibnamefont {Hensen}}, \bibinfo {author} {\bibfnamefont
  {T.}~\bibnamefont {Tanttu}}, \bibinfo {author} {\bibfnamefont {F.~E.}\
  \bibnamefont {Hudson}}, \bibinfo {author} {\bibfnamefont {K.~M.}\
  \bibnamefont {Itoh}}, \bibinfo {author} {\bibfnamefont {A.}~\bibnamefont
  {Laucht}}, \bibinfo {author} {\bibfnamefont {A.}~\bibnamefont {Morello}}, \
  and\ \bibinfo {author} {\bibfnamefont {A.~S.}\ \bibnamefont {Dzurak}},\
  }\href {\doibase 10.1103/PhysRevApplied.10.044017} {\bibfield  {journal}
  {\bibinfo  {journal} {Physical Review Applied}\ }\textbf {\bibinfo {volume}
  {10}},\ \bibinfo {pages} {044017} (\bibinfo {year} {2018})}\BibitemShut
  {NoStop}%
\bibitem [{\citenamefont {Itoh}\ and\ \citenamefont
  {Watanabe}(2014)}]{Itoh2014}%
  \BibitemOpen
  \bibfield  {author} {\bibinfo {author} {\bibfnamefont {K.~M.}\ \bibnamefont
  {Itoh}}\ and\ \bibinfo {author} {\bibfnamefont {H.}~\bibnamefont
  {Watanabe}},\ }\href {\doibase 10.1557/mrc.2014.32} {\bibfield  {journal}
  {\bibinfo  {journal} {MRS Communications}\ }\textbf {\bibinfo {volume} {4}},\
  \bibinfo {pages} {143} (\bibinfo {year} {2014})}\BibitemShut {NoStop}%
\bibitem [{\citenamefont {Yang}\ \emph {et~al.}(2019)\citenamefont {Yang},
  \citenamefont {Chan}, \citenamefont {Harper}, \citenamefont {Huang},
  \citenamefont {Evans}, \citenamefont {Hwang}, \citenamefont {Hensen},
  \citenamefont {Laucht}, \citenamefont {Tanttu}, \citenamefont {Hudson},
  \citenamefont {Flammia}, \citenamefont {Itoh}, \citenamefont {Morello},
  \citenamefont {Bartlett},\ and\ \citenamefont {Dzurak}}]{Yang2019}%
  \BibitemOpen
  \bibfield  {author} {\bibinfo {author} {\bibfnamefont {C.~H.}\ \bibnamefont
  {Yang}}, \bibinfo {author} {\bibfnamefont {K.~W.}\ \bibnamefont {Chan}},
  \bibinfo {author} {\bibfnamefont {R.}~\bibnamefont {Harper}}, \bibinfo
  {author} {\bibfnamefont {W.}~\bibnamefont {Huang}}, \bibinfo {author}
  {\bibfnamefont {T.}~\bibnamefont {Evans}}, \bibinfo {author} {\bibfnamefont
  {J.~C.~C.}\ \bibnamefont {Hwang}}, \bibinfo {author} {\bibfnamefont
  {B.}~\bibnamefont {Hensen}}, \bibinfo {author} {\bibfnamefont
  {A.}~\bibnamefont {Laucht}}, \bibinfo {author} {\bibfnamefont
  {T.}~\bibnamefont {Tanttu}}, \bibinfo {author} {\bibfnamefont {F.~E.}\
  \bibnamefont {Hudson}}, \bibinfo {author} {\bibfnamefont {S.~T.}\
  \bibnamefont {Flammia}}, \bibinfo {author} {\bibfnamefont {K.~M.}\
  \bibnamefont {Itoh}}, \bibinfo {author} {\bibfnamefont {A.}~\bibnamefont
  {Morello}}, \bibinfo {author} {\bibfnamefont {S.~D.}\ \bibnamefont
  {Bartlett}}, \ and\ \bibinfo {author} {\bibfnamefont {A.~S.}\ \bibnamefont
  {Dzurak}},\ }\href {\doibase 10.1038/s41928-019-0234-1} {\bibfield  {journal}
  {\bibinfo  {journal} {Nature Electronics}\ }\textbf {\bibinfo {volume} {2}},\
  \bibinfo {pages} {151} (\bibinfo {year} {2019})}\BibitemShut {NoStop}%
\bibitem [{\citenamefont {Vepsäläinen}\ \emph {et~al.}(2022)\citenamefont
  {Vepsäläinen}, \citenamefont {Winik}, \citenamefont {Karamlou},
  \citenamefont {Braumüller}, \citenamefont {Paolo}, \citenamefont {Sung},
  \citenamefont {Kannan}, \citenamefont {Kjaergaard}, \citenamefont {Kim},
  \citenamefont {Melville}, \citenamefont {Niedzielski}, \citenamefont {Yoder},
  \citenamefont {Gustavsson},\ and\ \citenamefont {Oliver}}]{Vep2022}%
  \BibitemOpen
  \bibfield  {author} {\bibinfo {author} {\bibfnamefont {A.}~\bibnamefont
  {Vepsäläinen}}, \bibinfo {author} {\bibfnamefont {R.}~\bibnamefont
  {Winik}}, \bibinfo {author} {\bibfnamefont {A.~H.}\ \bibnamefont {Karamlou}},
  \bibinfo {author} {\bibfnamefont {J.}~\bibnamefont {Braumüller}}, \bibinfo
  {author} {\bibfnamefont {A.~D.}\ \bibnamefont {Paolo}}, \bibinfo {author}
  {\bibfnamefont {Y.}~\bibnamefont {Sung}}, \bibinfo {author} {\bibfnamefont
  {B.}~\bibnamefont {Kannan}}, \bibinfo {author} {\bibfnamefont
  {M.}~\bibnamefont {Kjaergaard}}, \bibinfo {author} {\bibfnamefont {D.~K.}\
  \bibnamefont {Kim}}, \bibinfo {author} {\bibfnamefont {A.~J.}\ \bibnamefont
  {Melville}}, \bibinfo {author} {\bibfnamefont {B.~M.}\ \bibnamefont
  {Niedzielski}}, \bibinfo {author} {\bibfnamefont {J.~L.}\ \bibnamefont
  {Yoder}}, \bibinfo {author} {\bibfnamefont {S.}~\bibnamefont {Gustavsson}}, \
  and\ \bibinfo {author} {\bibfnamefont {W.~D.}\ \bibnamefont {Oliver}},\
  }\href {\doibase 10.1038/s41467-022-29287-4} {\bibfield  {journal} {\bibinfo
  {journal} {Nature Communications}\ }\textbf {\bibinfo {volume} {13}},\
  \bibinfo {pages} {1932} (\bibinfo {year} {2022})}\BibitemShut {NoStop}%
\bibitem [{\citenamefont {Philips}\ \emph {et~al.}(2022)\citenamefont
  {Philips}, \citenamefont {Madzik}, \citenamefont {Amitonov}, \citenamefont
  {de~Snoo}, \citenamefont {Russ}, \citenamefont {Kalhor}, \citenamefont
  {Volk}, \citenamefont {Lawrie}, \citenamefont {Brousse}, \citenamefont
  {Tryputen}, \citenamefont {Wuetz}, \citenamefont {Sammak}, \citenamefont
  {Veldhorst}, \citenamefont {Scappucci},\ and\ \citenamefont
  {Vandersypen}}]{philips2022}%
  \BibitemOpen
  \bibfield  {author} {\bibinfo {author} {\bibfnamefont {S.~G.~J.}\
  \bibnamefont {Philips}}, \bibinfo {author} {\bibfnamefont {M.~T.}\
  \bibnamefont {Madzik}}, \bibinfo {author} {\bibfnamefont {S.~V.}\
  \bibnamefont {Amitonov}}, \bibinfo {author} {\bibfnamefont {S.~L.}\
  \bibnamefont {de~Snoo}}, \bibinfo {author} {\bibfnamefont {M.}~\bibnamefont
  {Russ}}, \bibinfo {author} {\bibfnamefont {N.}~\bibnamefont {Kalhor}},
  \bibinfo {author} {\bibfnamefont {C.}~\bibnamefont {Volk}}, \bibinfo {author}
  {\bibfnamefont {W.~I.~L.}\ \bibnamefont {Lawrie}}, \bibinfo {author}
  {\bibfnamefont {D.}~\bibnamefont {Brousse}}, \bibinfo {author} {\bibfnamefont
  {L.}~\bibnamefont {Tryputen}}, \bibinfo {author} {\bibfnamefont {B.~P.}\
  \bibnamefont {Wuetz}}, \bibinfo {author} {\bibfnamefont {A.}~\bibnamefont
  {Sammak}}, \bibinfo {author} {\bibfnamefont {M.}~\bibnamefont {Veldhorst}},
  \bibinfo {author} {\bibfnamefont {G.}~\bibnamefont {Scappucci}}, \ and\
  \bibinfo {author} {\bibfnamefont {L.~M.~K.}\ \bibnamefont {Vandersypen}},\
  }\href {\doibase 10.1038/s41586-022-05117-x} {\bibfield  {journal} {\bibinfo
  {journal} {Nature}\ }\textbf {\bibinfo {volume} {609}},\ \bibinfo {pages}
  {919} (\bibinfo {year} {2022})}\BibitemShut {NoStop}%
\bibitem [{\citenamefont {Paladino}\ \emph {et~al.}(2014)\citenamefont
  {Paladino}, \citenamefont {Galperin}, \citenamefont {Falci},\ and\
  \citenamefont {Altshuler}}]{Paladino2014}%
  \BibitemOpen
  \bibfield  {author} {\bibinfo {author} {\bibfnamefont {E.}~\bibnamefont
  {Paladino}}, \bibinfo {author} {\bibfnamefont {Y.~M.}\ \bibnamefont
  {Galperin}}, \bibinfo {author} {\bibfnamefont {G.}~\bibnamefont {Falci}}, \
  and\ \bibinfo {author} {\bibfnamefont {B.~L.}\ \bibnamefont {Altshuler}},\
  }\href {\doibase 10.1103/RevModPhys.86.361} {\bibfield  {journal} {\bibinfo
  {journal} {Reviews of Modern Physics}\ }\textbf {\bibinfo {volume} {86}},\
  \bibinfo {pages} {361} (\bibinfo {year} {2014})}\BibitemShut {NoStop}%
\bibitem [{\citenamefont {Baur}\ \emph {et~al.}(2009)\citenamefont {Baur},
  \citenamefont {Filipp}, \citenamefont {Bianchetti}, \citenamefont {Fink},
  \citenamefont {Göppl}, \citenamefont {Steffen}, \citenamefont {Leek},
  \citenamefont {Blais},\ and\ \citenamefont {Wallraff}}]{Baur2009}%
  \BibitemOpen
  \bibfield  {author} {\bibinfo {author} {\bibfnamefont {M.}~\bibnamefont
  {Baur}}, \bibinfo {author} {\bibfnamefont {S.}~\bibnamefont {Filipp}},
  \bibinfo {author} {\bibfnamefont {R.}~\bibnamefont {Bianchetti}}, \bibinfo
  {author} {\bibfnamefont {J.~M.}\ \bibnamefont {Fink}}, \bibinfo {author}
  {\bibfnamefont {M.}~\bibnamefont {Göppl}}, \bibinfo {author} {\bibfnamefont
  {L.}~\bibnamefont {Steffen}}, \bibinfo {author} {\bibfnamefont {P.~J.}\
  \bibnamefont {Leek}}, \bibinfo {author} {\bibfnamefont {A.}~\bibnamefont
  {Blais}}, \ and\ \bibinfo {author} {\bibfnamefont {A.}~\bibnamefont
  {Wallraff}},\ }\href {\doibase 10.1103/PhysRevLett.102.243602} {\bibfield
  {journal} {\bibinfo  {journal} {Physical Review Letters}\ }\textbf {\bibinfo
  {volume} {102}},\ \bibinfo {pages} {243602} (\bibinfo {year}
  {2009})}\BibitemShut {NoStop}%
\bibitem [{\citenamefont {Laucht}\ \emph {et~al.}(2017)\citenamefont {Laucht},
  \citenamefont {Kalra}, \citenamefont {Simmons}, \citenamefont {Dehollain},
  \citenamefont {Muhonen}, \citenamefont {Mohiyaddin}, \citenamefont {Freer},
  \citenamefont {Hudson}, \citenamefont {Itoh}, \citenamefont {Jamieson},
  \citenamefont {McCallum}, \citenamefont {Dzurak},\ and\ \citenamefont
  {Morello}}]{laucht2017dressed}%
  \BibitemOpen
  \bibfield  {author} {\bibinfo {author} {\bibfnamefont {A.}~\bibnamefont
  {Laucht}}, \bibinfo {author} {\bibfnamefont {R.}~\bibnamefont {Kalra}},
  \bibinfo {author} {\bibfnamefont {S.}~\bibnamefont {Simmons}}, \bibinfo
  {author} {\bibfnamefont {J.~P.}\ \bibnamefont {Dehollain}}, \bibinfo {author}
  {\bibfnamefont {J.~T.}\ \bibnamefont {Muhonen}}, \bibinfo {author}
  {\bibfnamefont {F.~A.}\ \bibnamefont {Mohiyaddin}}, \bibinfo {author}
  {\bibfnamefont {S.}~\bibnamefont {Freer}}, \bibinfo {author} {\bibfnamefont
  {F.~E.}\ \bibnamefont {Hudson}}, \bibinfo {author} {\bibfnamefont {K.~M.}\
  \bibnamefont {Itoh}}, \bibinfo {author} {\bibfnamefont {D.~N.}\ \bibnamefont
  {Jamieson}}, \bibinfo {author} {\bibfnamefont {J.~C.}\ \bibnamefont
  {McCallum}}, \bibinfo {author} {\bibfnamefont {A.~S.}\ \bibnamefont
  {Dzurak}}, \ and\ \bibinfo {author} {\bibfnamefont {A.}~\bibnamefont
  {Morello}},\ }\href {\doibase 10.1038/nnano.2016.178} {\bibfield  {journal}
  {\bibinfo  {journal} {Nature Nanotechnology}\ }\textbf {\bibinfo {volume}
  {12}},\ \bibinfo {pages} {61} (\bibinfo {year} {2017})}\BibitemShut {NoStop}%
\bibitem [{\citenamefont {Seedhouse}\ \emph {et~al.}(2021)\citenamefont
  {Seedhouse}, \citenamefont {Hansen}, \citenamefont {Laucht}, \citenamefont
  {Yang}, \citenamefont {Dzurak},\ and\ \citenamefont
  {Saraiva}}]{Seedhouse2021}%
  \BibitemOpen
  \bibfield  {author} {\bibinfo {author} {\bibfnamefont {A.~E.}\ \bibnamefont
  {Seedhouse}}, \bibinfo {author} {\bibfnamefont {I.}~\bibnamefont {Hansen}},
  \bibinfo {author} {\bibfnamefont {A.}~\bibnamefont {Laucht}}, \bibinfo
  {author} {\bibfnamefont {C.~H.}\ \bibnamefont {Yang}}, \bibinfo {author}
  {\bibfnamefont {A.~S.}\ \bibnamefont {Dzurak}}, \ and\ \bibinfo {author}
  {\bibfnamefont {A.}~\bibnamefont {Saraiva}},\ }\href {\doibase
  10.1103/PhysRevB.104.235411} {\bibfield  {journal} {\bibinfo  {journal}
  {Phys. Rev. B}\ }\textbf {\bibinfo {volume} {104}},\ \bibinfo {pages}
  {235411} (\bibinfo {year} {2021})}\BibitemShut {NoStop}%
\bibitem [{\citenamefont {Hansen}\ \emph {et~al.}(2021)\citenamefont {Hansen},
  \citenamefont {Seedhouse}, \citenamefont {Saraiva}, \citenamefont {Laucht},
  \citenamefont {Dzurak},\ and\ \citenamefont {Yang}}]{Hansen2021}%
  \BibitemOpen
  \bibfield  {author} {\bibinfo {author} {\bibfnamefont {I.}~\bibnamefont
  {Hansen}}, \bibinfo {author} {\bibfnamefont {A.~E.}\ \bibnamefont
  {Seedhouse}}, \bibinfo {author} {\bibfnamefont {A.}~\bibnamefont {Saraiva}},
  \bibinfo {author} {\bibfnamefont {A.}~\bibnamefont {Laucht}}, \bibinfo
  {author} {\bibfnamefont {A.~S.}\ \bibnamefont {Dzurak}}, \ and\ \bibinfo
  {author} {\bibfnamefont {C.~H.}\ \bibnamefont {Yang}},\ }\href {\doibase
  10.1103/PhysRevA.104.062415} {\bibfield  {journal} {\bibinfo  {journal}
  {Phys. Rev. A}\ }\textbf {\bibinfo {volume} {104}},\ \bibinfo {pages}
  {062415} (\bibinfo {year} {2021})}\BibitemShut {NoStop}%
\bibitem [{\citenamefont {Kane}(1998)}]{kane1998}%
  \BibitemOpen
  \bibfield  {author} {\bibinfo {author} {\bibfnamefont {B.~E.}\ \bibnamefont
  {Kane}},\ }\href {\doibase 10.1038/30156} {\bibfield  {journal} {\bibinfo
  {journal} {Nature}\ }\textbf {\bibinfo {volume} {393}},\ \bibinfo {pages}
  {133} (\bibinfo {year} {1998})}\BibitemShut {NoStop}%
\bibitem [{\citenamefont {Zeng}\ \emph {et~al.}(2019)\citenamefont {Zeng},
  \citenamefont {Yang}, \citenamefont {Dzurak},\ and\ \citenamefont
  {Barnes}}]{Zeng2019}%
  \BibitemOpen
  \bibfield  {author} {\bibinfo {author} {\bibfnamefont {J.}~\bibnamefont
  {Zeng}}, \bibinfo {author} {\bibfnamefont {C.~H.}\ \bibnamefont {Yang}},
  \bibinfo {author} {\bibfnamefont {A.~S.}\ \bibnamefont {Dzurak}}, \ and\
  \bibinfo {author} {\bibfnamefont {E.}~\bibnamefont {Barnes}},\ }\href
  {\doibase 10.1103/PhysRevA.99.052321} {\bibfield  {journal} {\bibinfo
  {journal} {Phys. Rev. A}\ }\textbf {\bibinfo {volume} {99}},\ \bibinfo
  {pages} {052321} (\bibinfo {year} {2019})}\BibitemShut {NoStop}%
\bibitem [{\citenamefont {Barnes}\ \emph {et~al.}(2022)\citenamefont {Barnes},
  \citenamefont {Calderon-Vargas}, \citenamefont {Dong}, \citenamefont {Li},
  \citenamefont {Zeng},\ and\ \citenamefont {Zhuang}}]{Barnes_2022}%
  \BibitemOpen
  \bibfield  {author} {\bibinfo {author} {\bibfnamefont {E.}~\bibnamefont
  {Barnes}}, \bibinfo {author} {\bibfnamefont {F.~A.}\ \bibnamefont
  {Calderon-Vargas}}, \bibinfo {author} {\bibfnamefont {W.}~\bibnamefont
  {Dong}}, \bibinfo {author} {\bibfnamefont {B.}~\bibnamefont {Li}}, \bibinfo
  {author} {\bibfnamefont {J.}~\bibnamefont {Zeng}}, \ and\ \bibinfo {author}
  {\bibfnamefont {F.}~\bibnamefont {Zhuang}},\ }\href {\doibase
  10.1088/2058-9565/ac4421} {\bibfield  {journal} {\bibinfo  {journal} {Quantum
  Science and Technology}\ }\textbf {\bibinfo {volume} {7}},\ \bibinfo {pages}
  {023001} (\bibinfo {year} {2022})}\BibitemShut {NoStop}%
\bibitem [{\citenamefont {Green}\ \emph {et~al.}(2013)\citenamefont {Green},
  \citenamefont {Sastrawan}, \citenamefont {Uys},\ and\ \citenamefont
  {Biercuk}}]{Green2013}%
  \BibitemOpen
  \bibfield  {author} {\bibinfo {author} {\bibfnamefont {T.~J.}\ \bibnamefont
  {Green}}, \bibinfo {author} {\bibfnamefont {J.}~\bibnamefont {Sastrawan}},
  \bibinfo {author} {\bibfnamefont {H.}~\bibnamefont {Uys}}, \ and\ \bibinfo
  {author} {\bibfnamefont {M.~J.}\ \bibnamefont {Biercuk}},\ }\href {\doibase
  10.1088/1367-2630/15/9/095004} {\bibfield  {journal} {\bibinfo  {journal}
  {New Journal of Physics}\ }\textbf {\bibinfo {volume} {15}},\ \bibinfo
  {pages} {095004} (\bibinfo {year} {2013})}\BibitemShut {NoStop}%
\bibitem [{\citenamefont {Hangleiter}\ \emph {et~al.}(2021)\citenamefont
  {Hangleiter}, \citenamefont {Cerfontaine},\ and\ \citenamefont
  {Bluhm}}]{Hangleiter2021}%
  \BibitemOpen
  \bibfield  {author} {\bibinfo {author} {\bibfnamefont {T.}~\bibnamefont
  {Hangleiter}}, \bibinfo {author} {\bibfnamefont {P.}~\bibnamefont
  {Cerfontaine}}, \ and\ \bibinfo {author} {\bibfnamefont {H.}~\bibnamefont
  {Bluhm}},\ }\href {\doibase 10.1103/PhysRevResearch.3.043047} {\bibfield
  {journal} {\bibinfo  {journal} {Phys. Rev. Research}\ }\textbf {\bibinfo
  {volume} {3}},\ \bibinfo {pages} {043047} (\bibinfo {year}
  {2021})}\BibitemShut {NoStop}%
\bibitem [{\citenamefont {Cerfontaine}\ \emph {et~al.}(2021)\citenamefont
  {Cerfontaine}, \citenamefont {Hangleiter},\ and\ \citenamefont
  {Bluhm}}]{Cerfontaine2021}%
  \BibitemOpen
  \bibfield  {author} {\bibinfo {author} {\bibfnamefont {P.}~\bibnamefont
  {Cerfontaine}}, \bibinfo {author} {\bibfnamefont {T.}~\bibnamefont
  {Hangleiter}}, \ and\ \bibinfo {author} {\bibfnamefont {H.}~\bibnamefont
  {Bluhm}},\ }\href {\doibase 10.1103/PhysRevLett.127.170403} {\bibfield
  {journal} {\bibinfo  {journal} {Phys. Rev. Lett.}\ }\textbf {\bibinfo
  {volume} {127}},\ \bibinfo {pages} {170403} (\bibinfo {year}
  {2021})}\BibitemShut {NoStop}%
\bibitem [{\citenamefont {Hansen}\ \emph {et~al.}(2022)\citenamefont {Hansen},
  \citenamefont {Seedhouse}, \citenamefont {Chan}, \citenamefont {Hudson},
  \citenamefont {Itoh}, \citenamefont {Laucht}, \citenamefont {Saraiva},
  \citenamefont {Yang},\ and\ \citenamefont {Dzurak}}]{hansen2022}%
  \BibitemOpen
  \bibfield  {author} {\bibinfo {author} {\bibfnamefont {I.}~\bibnamefont
  {Hansen}}, \bibinfo {author} {\bibfnamefont {A.~E.}\ \bibnamefont
  {Seedhouse}}, \bibinfo {author} {\bibfnamefont {K.~W.}\ \bibnamefont {Chan}},
  \bibinfo {author} {\bibfnamefont {F.~E.}\ \bibnamefont {Hudson}}, \bibinfo
  {author} {\bibfnamefont {K.~M.}\ \bibnamefont {Itoh}}, \bibinfo {author}
  {\bibfnamefont {A.}~\bibnamefont {Laucht}}, \bibinfo {author} {\bibfnamefont
  {A.}~\bibnamefont {Saraiva}}, \bibinfo {author} {\bibfnamefont {C.~H.}\
  \bibnamefont {Yang}}, \ and\ \bibinfo {author} {\bibfnamefont {A.~S.}\
  \bibnamefont {Dzurak}},\ }\href {\doibase 10.1063/5.0096467} {\bibfield
  {journal} {\bibinfo  {journal} {Applied Physics Reviews}\ }\textbf {\bibinfo
  {volume} {9}},\ \bibinfo {pages} {031409} (\bibinfo {year}
  {2022})}\BibitemShut {NoStop}%
\bibitem [{\citenamefont {Biercuk}\ \emph {et~al.}(2009)\citenamefont
  {Biercuk}, \citenamefont {Uys}, \citenamefont {VanDevender}, \citenamefont
  {Shiga}, \citenamefont {Itano},\ and\ \citenamefont
  {Bollinger}}]{Biercuk2009}%
  \BibitemOpen
  \bibfield  {author} {\bibinfo {author} {\bibfnamefont {M.~J.}\ \bibnamefont
  {Biercuk}}, \bibinfo {author} {\bibfnamefont {H.}~\bibnamefont {Uys}},
  \bibinfo {author} {\bibfnamefont {A.~P.}\ \bibnamefont {VanDevender}},
  \bibinfo {author} {\bibfnamefont {N.}~\bibnamefont {Shiga}}, \bibinfo
  {author} {\bibfnamefont {W.~M.}\ \bibnamefont {Itano}}, \ and\ \bibinfo
  {author} {\bibfnamefont {J.~J.}\ \bibnamefont {Bollinger}},\ }\href {\doibase
  10.1038/nature07951} {\bibfield  {journal} {\bibinfo  {journal} {Nature}\
  }\textbf {\bibinfo {volume} {458}},\ \bibinfo {pages} {996} (\bibinfo {year}
  {2009})}\BibitemShut {NoStop}%
\bibitem [{\citenamefont {Biercuk}\ \emph {et~al.}(2011)\citenamefont
  {Biercuk}, \citenamefont {Doherty},\ and\ \citenamefont {Uys}}]{Biercuk2011}%
  \BibitemOpen
  \bibfield  {author} {\bibinfo {author} {\bibfnamefont {M.~J.}\ \bibnamefont
  {Biercuk}}, \bibinfo {author} {\bibfnamefont {A.~C.}\ \bibnamefont
  {Doherty}}, \ and\ \bibinfo {author} {\bibfnamefont {H.}~\bibnamefont
  {Uys}},\ }\href {\doibase 10.1088/0953-4075/44/15/154002} {\bibfield
  {journal} {\bibinfo  {journal} {Journal of Physics B: Atomic, Molecular and
  Optical Physics}\ }\textbf {\bibinfo {volume} {44}},\ \bibinfo {pages}
  {154002} (\bibinfo {year} {2011})}\BibitemShut {NoStop}%
\bibitem [{\citenamefont {Frey}\ \emph {et~al.}(2017)\citenamefont {Frey},
  \citenamefont {Mavadia}, \citenamefont {Norris}, \citenamefont {de~Ferranti},
  \citenamefont {Lucarelli}, \citenamefont {Viola},\ and\ \citenamefont
  {Biercuk}}]{Frey2017}%
  \BibitemOpen
  \bibfield  {author} {\bibinfo {author} {\bibfnamefont {V.~M.}\ \bibnamefont
  {Frey}}, \bibinfo {author} {\bibfnamefont {S.}~\bibnamefont {Mavadia}},
  \bibinfo {author} {\bibfnamefont {L.~M.}\ \bibnamefont {Norris}}, \bibinfo
  {author} {\bibfnamefont {W.}~\bibnamefont {de~Ferranti}}, \bibinfo {author}
  {\bibfnamefont {D.}~\bibnamefont {Lucarelli}}, \bibinfo {author}
  {\bibfnamefont {L.}~\bibnamefont {Viola}}, \ and\ \bibinfo {author}
  {\bibfnamefont {M.~J.}\ \bibnamefont {Biercuk}},\ }\href {\doibase
  10.1038/s41467-017-02298-2} {\bibfield  {journal} {\bibinfo  {journal}
  {Nature Communications}\ }\textbf {\bibinfo {volume} {8}},\ \bibinfo {pages}
  {2189} (\bibinfo {year} {2017})}\BibitemShut {NoStop}%
\bibitem [{\citenamefont {Kofman}\ and\ \citenamefont
  {Kurizki}(2001)}]{Kofman2001}%
  \BibitemOpen
  \bibfield  {author} {\bibinfo {author} {\bibfnamefont {A.~G.}\ \bibnamefont
  {Kofman}}\ and\ \bibinfo {author} {\bibfnamefont {G.}~\bibnamefont
  {Kurizki}},\ }\href {\doibase 10.1103/PhysRevLett.87.270405} {\bibfield
  {journal} {\bibinfo  {journal} {Phys. Rev. Lett.}\ }\textbf {\bibinfo
  {volume} {87}},\ \bibinfo {pages} {270405} (\bibinfo {year}
  {2001})}\BibitemShut {NoStop}%
\bibitem [{\citenamefont {Kofman}\ and\ \citenamefont
  {Kurizki}(2004)}]{Kofman2004}%
  \BibitemOpen
  \bibfield  {author} {\bibinfo {author} {\bibfnamefont {A.~G.}\ \bibnamefont
  {Kofman}}\ and\ \bibinfo {author} {\bibfnamefont {G.}~\bibnamefont
  {Kurizki}},\ }\href {\doibase 10.1103/PhysRevLett.93.130406} {\bibfield
  {journal} {\bibinfo  {journal} {Phys. Rev. Lett.}\ }\textbf {\bibinfo
  {volume} {93}},\ \bibinfo {pages} {130406} (\bibinfo {year}
  {2004})}\BibitemShut {NoStop}%
\bibitem [{\citenamefont {Arnal}\ \emph {et~al.}(2018)\citenamefont {Arnal},
  \citenamefont {Casas},\ and\ \citenamefont {Chiralt}}]{Arnal2018}%
  \BibitemOpen
  \bibfield  {author} {\bibinfo {author} {\bibfnamefont {A.}~\bibnamefont
  {Arnal}}, \bibinfo {author} {\bibfnamefont {F.}~\bibnamefont {Casas}}, \ and\
  \bibinfo {author} {\bibfnamefont {C.}~\bibnamefont {Chiralt}},\ }\href
  {\doibase 10.1088/2399-6528/aab291} {\bibfield  {journal} {\bibinfo
  {journal} {Journal of Physics Communications}\ }\textbf {\bibinfo {volume}
  {2}},\ \bibinfo {pages} {035024} (\bibinfo {year} {2018})}\BibitemShut
  {NoStop}%
\end{thebibliography}%

\end{document}